\documentclass[conference]{IEEEtran}
\IEEEoverridecommandlockouts

\usepackage{cite}
\usepackage{amsmath,amssymb,amsfonts}
\usepackage{algorithmic}
\usepackage{graphicx}
\usepackage{textcomp}
\usepackage{xcolor}
\usepackage{hyperref}

\newcommand{\sgc}[1]{}

\def\BibTeX{{\rm B\kern-.05em{\sc i\kern-.025em b}\kern-.08em
    T\kern-.1667em\lower.7ex\hbox{E}\kern-.125emX}}
\begin{document}

\title{Localization of Vibrotactile Stimuli on the Face\\
}

\author{\IEEEauthorblockN{Shivani Guptasarma}
\IEEEauthorblockA{\textit{Department of Mechanical Engineering} \\
\textit{Stanford University}\\
Stanford USA \\
shivanig@stanford.edu}
\and
\IEEEauthorblockN{Allison M. Okamura}
\IEEEauthorblockA{\textit{Department of Mechanical Engineering} \\
\textit{Stanford University}\\
Stanford, USA \\
aokamura@stanford.edu}
\and
\IEEEauthorblockN{Monroe Kennedy III}
\IEEEauthorblockA{\textit{Department of Mechanical Engineering} \\
\textit{Stanford University}\\
Stanford, USA \\
monroek@stanford.edu}
}

\maketitle

\begin{abstract}
The face remains relatively unexplored as a target region for haptic feedback, despite providing a considerable surface area consisting of highly sensitive skin. There are promising applications for facial haptic feedback, especially in cases of severe upper limb loss or spinal cord injury, where the face is typically less impacted than other body parts. Moreover, the neural representation of the face is adjacent to that of the hand, and phantom maps have been discovered between the fingertips and the cheeks. However, there is a dearth of compact devices for facial haptic feedback, and vibrotactile stimulation, a common modality of haptic feedback, has not been characterized for localization acuity on the face. We performed a localization experiment on the cheek, with an arrangement of off-the-shelf coin vibration motors. The study follows the methods of prior work studying other skin regionswhich participants attempt to identify the sites of discrete vibrotactile stimuli. We intend for our results to inform the future development of systems using vibrotactile feedback to convey information via the face.
\end{abstract}


\section{Introduction}
\label{sc:intro}
The hand and forearm are the most widely-studied regions for haptic feedback, but are not always available or suitable for applications where such feedback might be useful. Upper limb prostheses, for example, may be hugely improved by conveying haptic feedback from sensors on the prosthetic hand to the upper arm~\cite{dey2023}, but not when there is shoulder disarticulation, or forequarter amputation. Users of assistive devices with spinal cord injury might be able to control their devices better with haptic feedback (e.g.,~\cite{deo2021effects}), but might have lost sensitivity in the hand or forearm.   

Facial skin is usually available, with unaffected sensitivity, in both these scenarios; yet, it has not been used for sensory substitution, perhaps because designing any device for the face brings up unique concerns: it must not be bulky or uncomfortable, must not occlude the eyes, ears, nose or jaw, and must be socially acceptable. Design of haptic interfaces for the face typically focuses on Virtual Reality (VR) applications rather than everyday tasks. In these applications, the concerns mentioned above do not arise, as the goal is not to routinely convey information from elsewhere, but to simulate touch on the face itself, for a short period of time.  

We argue that there are several reasons to seriously consider the face as a target for haptic feedback, and to put efforts into designing facial haptic interfaces, wearable directly on the face, that account for its specific challenges. First, as mentioned, there exist applications for haptic feedback where the hand and forearm cannot be used. Second, there exist, already, several devices and accessories worn on the face, which have come to be regarded as ordinary, and there is reason to imagine that a well-designed haptic interface might do the same. Third, there has been fundamental research relating the neural representations of the fingertips and the face, and the plasticity of the brain regions responsible for sensation on the face and hands~\cite{ramachandran1998ThePO} -- which opens up immense potential for sensory substitution via the face in neural and myoelectric prostheses. 

In this work, we report a study on the localization accuracy of an array of coin vibration motors placed on the cheek using a skin-safe adhesive. We begin with a discussion of the design considerations and our approach towards choosing the vibrotactile modality. Then, we present the responses of~$10$ volunteers to vibrotactile stimulation of $12$~sites, actuated one at a time. We placed the actuators on the front and side of the cheek, taking care to cover the lower cheek, where phantom maps occur in some amputees~\cite{ramachandran1998ThePO}. Participants were asked to identify the location of each stimulus and to give free-form feedback about the interface. 

The contributions of this work are: (a) a discussion of the motivations and design considerations for facial haptic feedback for everyday applications, and (b) preliminary experimental results on the localization acuity of vibrotactile stimuli on the cheek. Based on our results, we propose future studies that can further inform the design of vibrotactile interfaces for the cheek.




\section{Related work}
This section consists of a review of the literature; first, on haptic feedback at locations other than the wrist and hand, and then, on previous studies of touch perception on the face.
\subsection{Alternative feedback sites}
\label{sc:elsewhere}
Feedback away from the hand and face has been mentioned in the literature in three different contexts: simulation of touch in VR, transmission of information when the hands are otherwise engaged, and sensory substitution for assistive devices. We summarize each of these below.

VR offers strong motivation for developing the ability to provide haptic feedback on areas other than the hand and forearm, to create more realistic experiences in simulated worlds (where objects may touch any part of the body). Since the circumference of the head extends beyond the face, in this work, devices modeled as headbands are treated separately from those designed for the ``face" -- while acknowledging that the forehead is an important site for haptic feedback, and is a subset of the skin area targeted by most head-worn devices. The head and torso have both been used for VR haptics~\cite{adilkhanoc2022review, kaul2016haptichead}. Previous work has also explored creative ways to provide a variety of stimuli to the face, using a manipulator arm mounted on the visor~\cite{wilberz2020}, ultrasonic arrays~\cite{shen2022, lan2024}, and multi-modal feedback, including vibrotactile and thermal sensations, on the contact surface between the skin and the visor~\cite{wolf2019face}. While effective for VR experiences, such devices are difficult to apply to the other potential use cases of facial haptics, as they are designed to be attached to a visor that covers the eyes, or to some other off-board mounting location.

Another reason to display information via haptic feedback away from the hands is so that the hands may remain unencumbered. The abdomen~\cite{cholewiakVibrotactileLocalizationAbdomen2004b}, front~\cite{vanerpAbsoluteLocalizationVibrotactile2008} and back~\cite{jones2008, jones2009} of the torso, whole torso~\cite{kim2023}, head~\cite{gilliland1994}, and earlobes~\cite{lee2019activearring} have all been studied for localization and pattern recognition performance, distributing vibrotactile actuators over their surface and quantifying recognition accuracy. Specific applications such as navigation guidance have been targeted through such methods, for example, for pilots~\cite{cholewiakVibrotactileLocalizationAbdomen2004b} and visually impaired users~\cite{katzschmann2018blind}.

Finally, as explained in Section~\ref{sc:intro}, the loss of a limb, or loss of sensation in the limb, can make it necessary to provide haptic feedback at the remaining available locations. It has been proposed to use the cheek for haptic feedback in robot teleoperation by possibly repurposing a pneumatically-actuated haptic feedback device designed for the forearm, to apply pressure to the cheek~\cite{guptasarma2024}. This work made a strong argument in favor of facial haptics, based on the high sensory innervation of facial skin~\cite{corniani2020tactile} and the proximity of neural regions mapping to the face and hands~\cite{leemhuis2022rethinking}. To these points, we add that the face is of particular interest for upper-limb amputees, as, in many cases, there exists a phantom map of the fingertips on the lower cheek~\cite{ramachandran1998ThePO}. However, to the best of our knowledge, there exist no experimental results relating to haptic feedback on the face with a technology \emph{mounted directly on the face} that is feasible to develop for everyday use. 

\subsection{Studies of facial touch sensitivity}
While facial haptics has not been studied comprehensively from the point of view of device design, there exists considerable literature on the sensitivity of the face to touch. The sensory innervation of the face is by the three branches of the trigeminal nerve, innervating the ophthalmic, maxillary, and mandibular areas respectively~\cite{gray1918}. These areas have been studied, both, from a neuroscience perspective, to identify the functional representation of the face in the somatosensory cortex~\cite{kikkert2023}, and from a clinical perspective, to restore sensation to healthy levels after surgical intervention~\cite{siemionow2011}.

Numerous studies report two-point discrimination tests, both static and dynamic, by various instruments, finding that resolution increases generally from superior to inferior regions, especially near the lips~\cite{costasNormalSensationHuman1994, fogacaEvaluationCutaneousSensibility2005, vriensExtensionNormalValues2009b}. In these works, moving points could be distinguished from each other at approximately a centimeter of separation on the cheek~\cite{fogacaEvaluationCutaneousSensibility2005}. However, as noted in previous works, these tests are not sufficient to characterize the response to vibratory stimuli~\cite{cholewiakVibrotactileLocalizationAbdomen2004b}. Studies on the detection of vibration on facial skin (obtained for the study of speech production)~\cite{barlow1987} have shown that displacement thresholds on the face exceed those on the fingertip, and are comparable to those on the forearm~\cite{morioka2008vibrotactile}. It was also shown that, on the face, these thresholds are not very sensitive to changes in frequency~\cite{barlow1987}.

From the point of view of device design, it is important to note that the aforementioned two-point discrimination tests are manually administered, without the ability to program stimuli into a device anchored to the face. Similarly, in works studying the correlation between facial touch and phantom fingertip sensation in amputees~\cite{ramachandran1998ThePO}, the skin was stroked by an experimenter (or held against a vibration source or wet object). One branch of research has sought to design specialized pneumatic devices, compatible with Magnetic Resonance (MR) environments, in order to provide more consistent vibrotactile stimuli for functional brain imaging studies~\cite{kikkert2023}. Being aimed at MR-compatibility, these designs are not suited for use in other settings; yet, as examples of device design for facial haptic feedback in the absence of a visor, they illuminate some important design considerations, such as the highly-variable curvature of faces which makes standardized devices difficult to attach. 
\section{Methods}
In this section, we first describe the design considerations behind to our decision to use the vibrotactile modality for this first experimental study, then report the materials and protocol for the experiment.

\subsection{Modality and mechanism}
For prosthetic and assistive applications, various types of information can be conveyed via haptics, including contact made or lost at different locations, the magnitude of normal forces, magnitude and direction of shear and torsion, as well as warmth, edges, and texture. Both discrete and continuous stimuli may be useful. In this study, we focus on discrete stimuli.

Haptic stimulation can be provided by skin deformation (poking~\cite{guptasarma2024}, pinching, twisting, or shearing~\cite{jayatilake2012stretch}), as well as vibration, thermal, or electrical stimulation. Because skin on the face is often compliant, the effective receptive field depends not only upon the mechanoreceptors stimulated, but also the size of the skin region that experiences displacements when a localized stimulus is delivered. The various stimuli can be capable of providing either continuous or discrete information, through switching on/off, or activating in spatiotemporal patterns (discrete) or the modulation of amplitude or frequency (continuous).

Skin deformation mechanisms have an advantage in that they often convey sensations through the same modality as originally intended (for example, mapping touch on the face to touch on the fingertip~\cite{ramachandran1998ThePO}), whereas vibration feedback is typically a form of sensory substitution. However, skin deformation mechanisms pose a considerable challenge because they require successful attachment and leverage on a body part that varies widely in shape, size, stiffness, texture, and skin health. Previous works describing such designs have faced this challenge by relying on custom face masks~\cite{kikkert2023}, since they were needed only inside MR scanners, or have attempted, with uncertain success, to use tight straps against bony protuberances~\cite{guptasarma2024}. Being made of soft materials, these were typically pneumatically driven, requiring noisy and bulky off-board compressors with a power supply. The prominent non-pneumatic designs have been the VR devices mentioned in Section~\ref{sc:elsewhere}, mounted on the headset visor.

In comparison to mechanisms for skin deformation feedback, vibrotactile actuators can be selected to be far more compact, such that even if the information that may be conveyed per actuator site is less (which is not known), more sites may be used over the available area. Being ubiquitous in smartphones, they have been developed to a point where they are widely available and economically accessible. Since their moving components are within their casing, they are also easy to mount on most surfaces.

In addition to mechanical considerations is a crucial logistical and social one: to be acceptable for continuous widespread use, a device designed for the face must not interfere with communication in any way, whether verbal or nonverbal, and must not impede mastication. It must not be physically uncomfortable or perceived as undesirable. In this context, it is encouraging to note that wearable items for the face and head are common in societies worldwide: including spectacles, monocles, hearing aids, earphones, headphones, headbands, over-the-ear microphones, earrings and noserings, among others. This list includes both active and passive devices, but with the active ones never undergoing large-scale external motions. Based on these observations, we hypothesize that, at present, vibrotactile feedback is the most promising choice for facial haptics, with actuators being relatively small, lightweight, and appearing far more passive (externally) than existing mechanisms for applying skin deformation. Thermal and electrical stimulation are also attractive for their potentially low profile configurations, but these actuators are less readily available.

\subsection{Vibrotactile actuators}
The three choices available for generating vibration are piezoelectric actuators, Eccentric Rotating Mass (ERM) motors and Linear Resonant Actuators (LRA). Piezoelectric actuators are taken out of consideration for this application by the high voltages required to drive them. The remaining two types of actuators are both suitable for haptic feedback. Frequency and amplitude are coupled in ERM motors, while LRAs offer greater control over the amplitude over a narrow range of frequency. LRAs are used with drivers to provide the required low-voltage alternating current, whereas ERM motors work with direct current (DC). As ERM is the more mature technology, these motors are more easily available and at lower cost. For this first study, we choose to use one stimulus at a fixed amplitude, frequency and pulse duration across all sites, hence, ERM motors are sufficient.

For this experiment, we used~$12$~Tatako~$12000$~RPM ($200$~Hz) coin motors of~$10$~mm diameter and~$3$~mm thickness, purchased for under~$0.70$~USD per motor. An appropriate frequency range was determined through a pilot test, concluding that very low frequencies were mechanically uncomfortable while very high frequencies created a shrill sound, undesirable so close to the ears. We also found that placing the motors on the nose occluded vision and created an unpleasant sensation (sometimes prompting sneezing), and that, on the forehead and tip of the chin, there was excessive transfer of vibration to the bone, which was also unpleasant. The decision was thus made to place the motors on the cheeks.

In order to achieve a medium frequency vibration, the motors (rated for~$3$~V~DC/$80$~mA) were actuated with a Pulse-Width Modulated signal with peak~$5$~V at a~$59\%$ duty cycle from the digital pins of an Arduino (Elegoo) Mega 2560 microcontroller board. The true amplitude of vibration for a face-mounted device (unlike when measuring thresholds with an externally-fixed device, e.g.,~\cite{barlow1987}) is dependent upon the mechanical response of the skin. The number of motors was selected as a balance between adequately covering the area of a single cheek and avoiding overburdening participants cognitively. The motors were placed at gaps of approximately~$10$~mm (one diameter, which is also the order of magnitude of the two-point discrimination threshold on the cheek~\cite{fogacaEvaluationCutaneousSensibility2005}), varying slightly with location. As vibration for prolonged periods can be uncomfortable, each individual stimulus lasted for~$200$~ms. The voltage across, and current through, the motors during operation were~$2.8$~V and $63$~mA respectively. 

\subsection{Protocol}
We recruited a total of~$10$~volunteers ($6$~female and~$4$~male) with no current injuries on the face and little to no facial hair on the cheeks. All participants gave informed consent. Participant~7 was cross-dominant, using his right hand for writing and fine motor control. The other nine participants were right-handed. All participants had uncorrected hearing and either uncorrected or fully-corrected vision.

\begin{figure}[htbp]
\centerline{\includegraphics[width=0.5\textwidth]{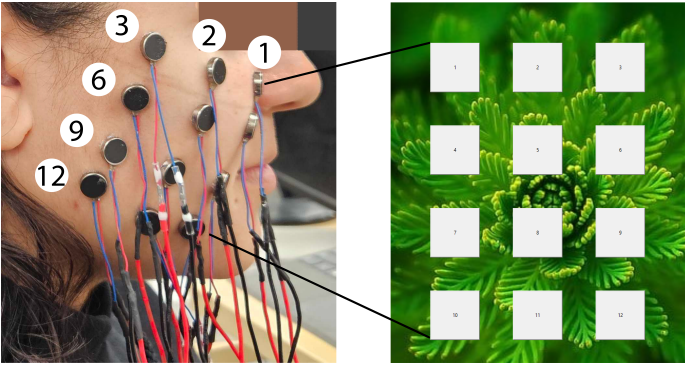}}
\caption{Left: the array of actuators on a participant's cheek; Right: the interface shown to participants for collecting responses.}
\label{fig:cheek}
\end{figure}

For consistency, we placed the motors in a~$4\times3$ grid drawn using a flexible stencil made by punching holes in a polyethylene sheet. The stencil was oriented with reference to a line drawn from the base of the right nostril to the top of the right ear, with the top row of motors falling above this line and the others below, as shown on the left in Fig.~\ref{fig:cheek}. To account for the different ways in which this standard grid might lie on participants' faces, photographs and measurements of facial dimensions were recorded for each participant, with consent.

Seating participants before a mirror and using wet wipes to prepare the skin, motors were attached to the right cheek, using a skin-safe adhesive (we used glue made for false eyelashes rather than for theater prosthetics, in the hope that it might be hypoallergenic for a greater part of the population). Adhesive tape was avoided, as were fabric masks, so as to minimize unnecessary stimulation of areas surrounding the attachment site, as well as to avoid the transmission of vibration between sites, through materials other than the cheek itself.

An array of buttons representing motor attachment sites (Fig.~\ref{fig:cheek}, right) was displayed on the touchscreen of a Microsoft Surface Pro~($2.50$~GHz processor,~$60$~Hz display). Due to the variation in face size and shape, and therefore, the placement relative to facial landmarks, a regular rectangular grid was shown on the screen, rather than a schematic of the face. A Python script running on the same device sent stimuli from a constrained random sequence, including each site~$3$ times, to the microcontroller. No two directly neighboring sites (within the same row or column) were actuated consecutively. Two such sessions were conducted without a break, placing the screen on the front-left or front-right of the participant respectively, so that the participant used their left and their right hand an equal number of times to touch the buttons.

As a calibration round prior to each of the two sessions, participants were asked to close their eyes and all sites from~$1$-$12$ were actuated while being verbally identified by the experimenter. During this time, participants were allowed to request the repetition of any stimuli until they were satisfied.

During the sessions, to ensure a consistent gaze direction, participants were asked to gaze at a certain point on the image displayed on the screen, while the buttons were hidden. The buttons reappeared~$0.5$~s after each stimulus was delivered, disappearing after each response. Each stimulus was delivered~$3$~s after the previous response.

Participants wore noise-canceling headphones, playing static, during sessions; however, it is impossible to isolate the effect of auditory cues on the behavior of hearing participants in such a study, since any vibrations on the cheek are transmitted to the bones, and can be heard internally.

Finally, each participant was given a brief verbal interview consisting of the following questions: (a) "Was the method of attachment uncomfortable?", and (b) "Was the sensation of vibration on the face unpleasant?". They were also asked if they had any other comments or observations. Selected insights from these comments are mentioned in Section~\ref{sc:results}.

\section{Results}
\label{sc:results}
\begin{figure*}[htbp]
\centerline{\includegraphics[width=\textwidth]{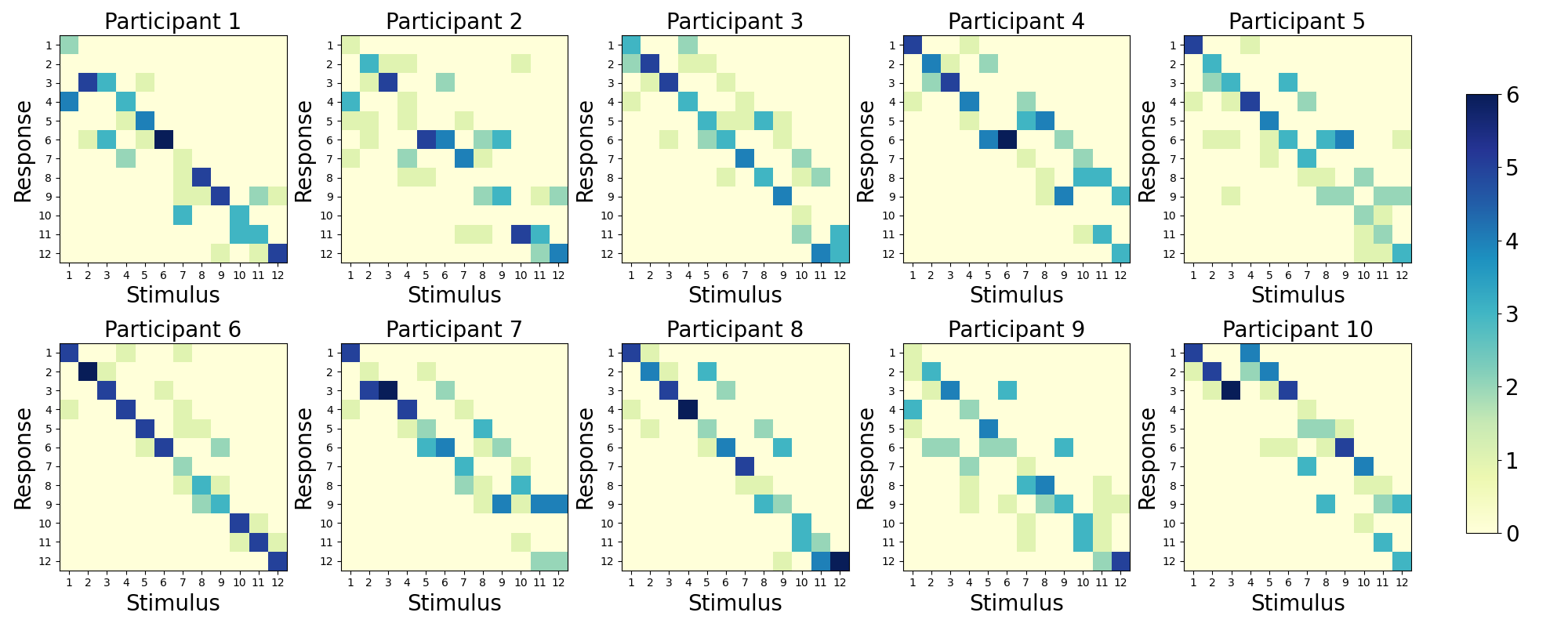}}
\caption{Confusion matrices showing correct responses along the diagonal, and misclassifications off the diagonal.}
\label{fig:conf}
\vspace{-1em}
\end{figure*}

\begin{figure}[htbp]
\centerline{\includegraphics[width=0.55\textwidth]{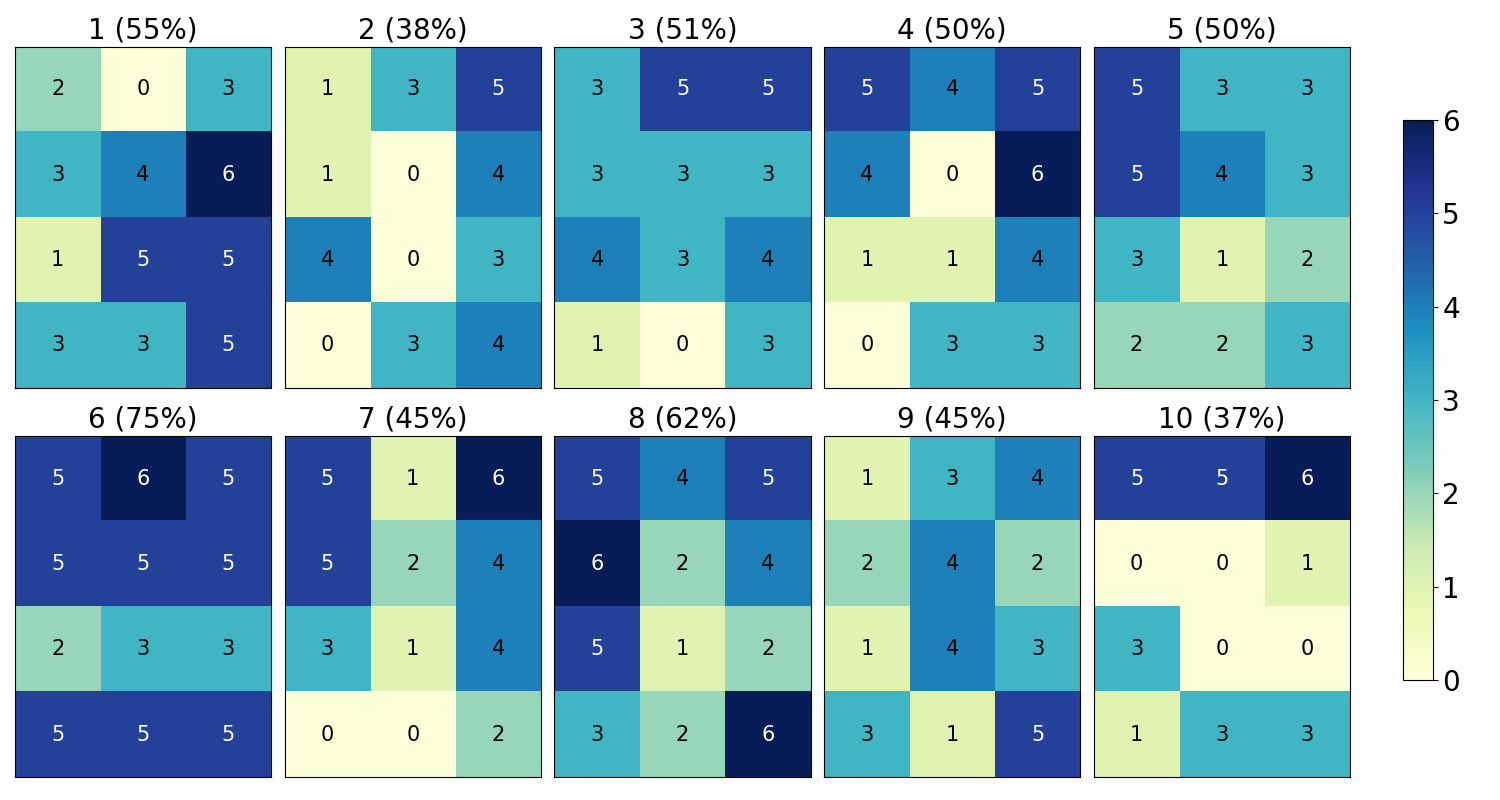}}
\caption{The number of correct responses per site, for each participant. L-R is medial-lateral in each grid. }
\label{fig:accgrid}
\vspace{-1em}
\end{figure}

Confusion matrices for all~$10$ participants are shown in Fig.~\ref{fig:conf}. Each stimulus was delivered a total of~$6$ times during the two sessions. 
The number of accurate responses across the~$12$ sites, and overall accuracy for each participant, are shown in Fig.~\ref{fig:accgrid}. Overall, it appears that sites near the extremes of the grid are the easiest to correctly identify. 

\begin{figure}[htbp]
\centerline{\includegraphics[width=0.6\textwidth]{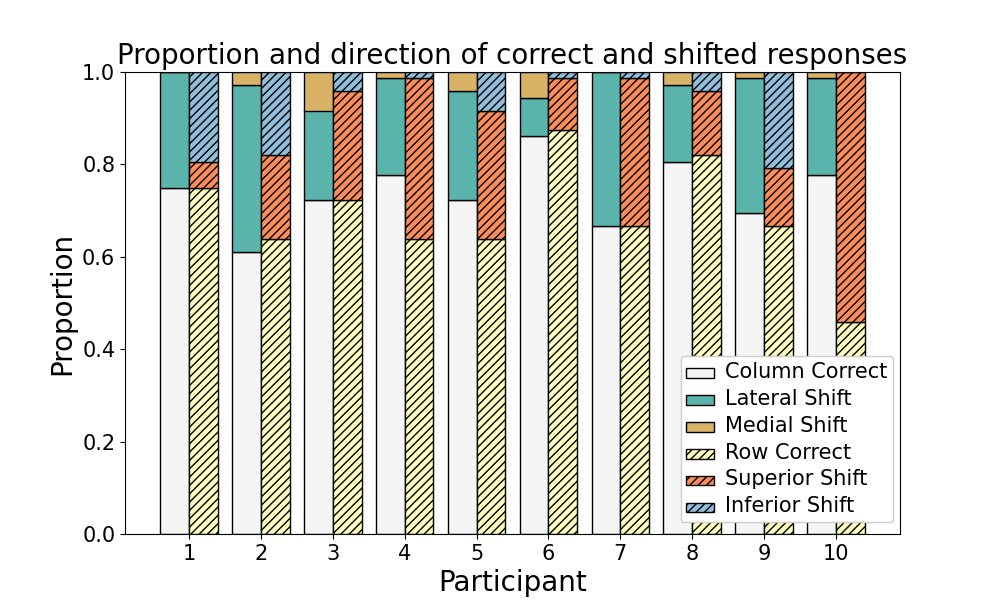}}
\caption{The directionality of errors by each participant; errors along columns and rows shown separately.}
\label{fig:shift}
\end{figure}

Aggregating errors at all sites, the shift from stimulus to response site was classified as medial or lateral along one axis, or inferior or superior along the other (Fig.~\ref{fig:shift}). Several participants perceived vibrations to be originating higher than their true origins (superior shift), albeit with a few exceptions. Meanwhile, there is a marked lateral shift (away from nose, towards ear) in the responses recorded from all participants.

\begin{figure}[htbp]
\centerline{\includegraphics[width=0.4\textwidth, trim={1em 0 1em 0}]{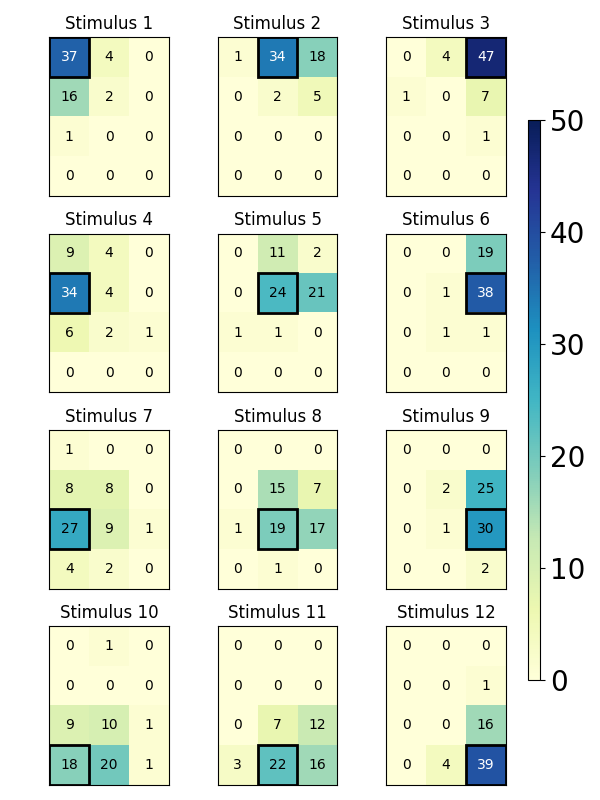}}
\caption{The frequencies of responses for each stimulus site, aggregated over participants. L-R is medial-lateral.}
\label{fig:hist}
\vspace{-1em}
\end{figure}

This lateral shift is also evident in Fig.\ref{fig:hist}, which shows the frequencies of responses given for each stimulus, aggregated over all participants. While stimuli close to the ear (3, 6, 9, and 12) and close to the nose and mouth (1, 4, and 7) were overwhelmingly localized to the correct column, those in the middle column were often misjudged to be closer to the ear. 

Subjective feedback from all participants confirmed that the attachment was not uncomfortable and vibration was not unpleasant. Two participants declared unprompted, during sessions, that it was a pleasant feeling. However, participants noted that if vibration was felt without warning (when the device was first connected), it was very startling. Participants reported being able to hear some vibrations through their bones, although opinions varied on whether this might have influenced their responses. A few participants mentioned that the softer part of the lower cheek, near the lip, was weighed down somewhat by the motors. Several participants remarked that the task was more difficult than expected, and that they wished the face were more sensitive.

\section{Discussion}
From the above results, it appears that while localization acuity on the cheek may be lower for vibrotactile stimuli than for light touch or pressure, localization is likely to be feasible with a smaller number of actuators placed at greater distances from each other, e.g., at the corners of the grid. 

A limitation of the setup was that some of the connecting wires could not be kept from touching the cheek (in real applications, cables may be tucked away behind the ear). While this was not uncomfortable for participants, it may have delivered additional vibrations at unintended locations. Motors placed very close to the eyes were also noticeable in the peripheral vision, which may or may not be acceptable in a real-world application.

The lateral shift found in the responses is reminiscent of similar phenomena in auditory localization~\cite{Lewald1996473} and in tactile localization on the waist~\cite{HO2007136}. In the present work, we did not study the relationship of head or eye position with this effect, but it is worthy of further study. 

Future work comparing the dominant and non-dominant side, and investigating the effect of head pose and eye gaze, is needed before vibrotactile stimulation can be adopted for haptic feedback applications on the face. It is necessary to identify the highest number of sites -- perhaps~$4$-$6$ -- that can be reliably distinguished. It is also of interest to experiment with the identification of spatiotemporal patterns, for example, simultaneous or sequential actuations at more than one site.

\errorcontextlines=99
\bibliography{main}{}
\bibliographystyle{IEEEtran}


\end{document}